\documentclass[aps,prl,twocolumn,groupedaddress,floats,showpacs]{revtex4}
\usepackage{latexsym}
\usepackage{dcolumn}
\usepackage[dvips]{graphicx}
\usepackage{amssymb}
\usepackage{graphics}
\usepackage{amsmath}
\usepackage{epsf}

\begin{document}

\author{S. Adam$^1$ and S. Cho$^2$ and M. S. Fuhrer$^2$ 
and S. Das Sarma$^{1,2}$}

\affiliation{$^1$Condensed Matter Theory Center,
Department of Physics, University of Maryland, College Park, MD
20742-4111, USA}

\affiliation{$^2$Center for Nanophysics and Advanced Materials,
Department of Physics, University of Maryland, College Park,
Maryland 20742-4111, USA}

\title{Density inhomogeneity driven percolation metal-insulator transition and 
dimensional crossover in graphene nanoribbons}
        
\date{\today}
\begin{abstract}
Transport in graphene nanoribbons with an energy gap in the spectrum
is considered in the presence of random charged impurity centers.  At
low carrier density, we predict and establish that the system exhibits
a density inhomogeneity driven two dimensional metal-insulator
transition that is in the percolation universality class.  For very
narrow graphene nanoribbons (with widths smaller than the disorder
induced length-scale), we predict that there should be a dimensional
crossover to the 1D percolation universality class with observable
signatures in the transport gap.  In addition, there should be a
crossover to the Boltzmann transport regime at high carrier densities.
The measured conductivity exponent and the critical density are
consistent with this percolation transition scenario.
\end{abstract}
\maketitle

One of the remarkable experimental findings of the past two
years, creating a great deal of activity and controversy, has been the
observation that the carrier density dependent dc conductivity of
gated 2D graphene layers, while being approximately linear in density
at high gate voltage,
becomes a non-universal constant over a finite gate voltage range
$\Delta V_g$ around the charge neutral Dirac point.  While this
conductivity minimum plateau formation around the charge neutrality
point in 2D graphene is experimentally well
established~\cite{kn:novoselov2004}, the actual values of $\sigma_{\rm
min} \sim 2~e^2/h - 20~e^2/h$ and $\Delta V_g \sim 1~V - 15~V$ are
non-universal and depend on the sample
quality~\cite{kn:tan2007,kn:chen2008}.  The minimum conductivity
plateau in graphene has been theoretically
explained~\cite{kn:hwang2006c,kn:adam2007a} to arise from the
invariable presence of unintentional random charged impurities at (or
near) the graphene-substrate interface which lead to inhomogeneous
electron-hole puddle formation in the low gate voltage
regime~\cite{kn:hwang2006c,kn:adam2007a,kn:rossi2008}.  We note that
distortions of the graphene membrane and quenched ripples can also
give rise to density
inhomogeneities~\cite{kn:brey2008}, and there
have been recent theories studying the effect of ripples on graphene
conductivity~\cite{kn:katsnelson2008}.
While we focus here on charged impurity induced inhomogeneities, many
of our conclusions are only sensitive to the existence of the
inhomogeneous density landscape (i.e. electron-hole ``puddles''), and
these do not distinguish between mechanisms (e.g. impurities, ripples)
producing these puddles.  Since graphene is a 2D semimetal (or more
appropriately, a {\it zero-gap} 2D chiral semiconductor with
electron-hole bands touching each other linearly at the charge neutral
Dirac point), the conductivity becomes an approximate constant when
the gate voltage induced chemical potential is pinned in this
electron-hole puddle region around the Dirac point.  This
inhomogeneous electron-hole puddle based theoretical
understanding of the graphene
minimum conductivity plateau formation leads immediately to an
important fundamental question: Are there situations where this
inhomogeneous puddle picture leads to a graphene 2D metal-insulator
transition (2D MIT) as is known~\cite{kn:dassarma2005b,kn:manfra2007}
to occur in 2D semiconductor systems?

We show in this Letter that indeed, as a direct consequence of the
inhomogeneous puddle formation in graphene, the system will manifest a
2D MIT, which is precisely in the same universality class as the
corresponding 2D MIT in electron~\cite{kn:dassarma2005b} and
hole~\cite{kn:manfra2007} GaAs systems, {\it provided that there is an
energy gap separating the graphene electron and hole bands}.  The
fundamental physics here is that of percolation -- for usual 2D
zero-gap graphene, percolation through the puddles is allowed at all
gate voltages, occurring either through the electron puddles or the
hole puddles (or through both~\cite{kn:cheianov2007}), since one or
the other is {\it always} percolating.  If there is a gap, however,
then there should be a percolation-driven 2D MIT in graphene exactly
as found~\cite{kn:dassarma2005b,kn:manfra2007} in 2D GaAs based
semiconductor structures.

The easiest way to introduce an energy gap in graphene, which would
then immediately lead to a percolation-induced transport gap (i.e. two
separate 2D MIT transitions for electrons and holes), is to consider
graphene {\it nanoribbons} instead of bulk 2D graphene.  In this
Letter, we predict and confirm experimentally that graphene
nanoribbons exhibit a 2D MIT in the low carrier density regime as a
function of the applied gate voltage and that this MIT is in the
percolation universality class; furthermore, we predict theoretically
that as ribbons become very narrow, there should be a dimensional
crossover to the 1D universality, implying that the observed transport
gap would tend to infinity as the ribbon width goes to zero (or in
practice, becomes smaller than the typical size of the puddles),
reflecting the 1D percolation universality where metallic conduction
is completely suppressed.  We speculate that such a 2D-1D crossover
may have been observed in recent
experiments~\cite{kn:han2007,kn:li2008}, but more quantitative work
and more data would be necessary to establish this prediction.

\begin{figure}
\bigskip
\epsfxsize=0.8\hsize
\hspace{0.0\hsize}
\epsffile{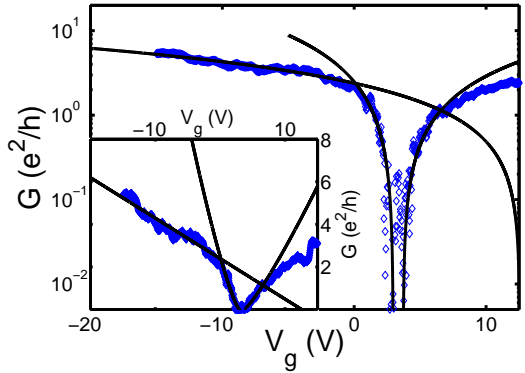}
\caption{\label{Fig:FuhrerA} Evidence of percolation
driven metal insulator transition in a graphene nanoribbon.  Main
panel shows graphene ribbon conductance as a function of gate voltage.
Best fits at low density to Eq.~\protect{\ref{Eq:Percolation}} give
for electrons $A = 1.485, n_c = 26.7485 \times 10^{10}~{\rm cm}^{-2},
\delta^e = 1.3 \pm 0.2$ and for holes $A = 1.755, n_c = - 18.5 \times
10^{10}~{\rm cm}^{-2}, \delta^h = 1.3 \pm 0.1$.  Best fit at high
density to Eq.~\protect{\ref{Eq:Boltzmann}} gives $n_{\rm imp} = 22
\times 10^{10}~{\rm cm}^{-2}$.  Inset shows the same data in a linear
scale, where even by eye the transition from high density Boltzmann
behavior to the low density percolation transport is visible.}
\end{figure}

\begin{figure}
\bigskip
\epsfxsize=0.7\hsize
\hspace{0.0\hsize}
\epsffile{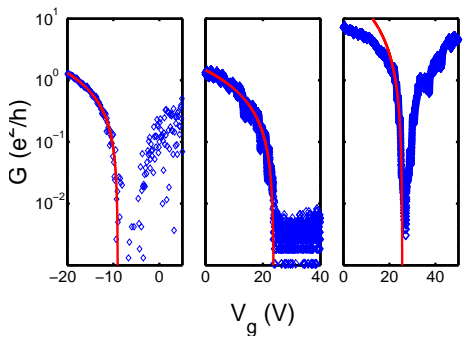}
\caption{\label{Fig:PanelFigure} Percolation 
driven metal-insulator transition in three additional graphene
samples.  The left panel shows a naturally occurring graphene
nanoribbon with dimensions $W \approx 200~{\rm nm}$ and $L \approx
11~{\rm \mu m }$ and has a critical exponent $\delta^h = 1.2 \pm 0.2$.
The center panel is the $W=24~{\rm nm}$ sample reported in
Ref.~\protect{\cite{kn:han2007}} which has a critical exponent of
$\delta^h = 1.3 \pm 0.1$, and the right panel is the $W=49~{\rm nm}$
sample reported in Ref.~\protect{\cite{kn:han2007}} which has a
critical exponent of $\delta^h = 1.6 \pm 0.3$.}
\end{figure}

In Fig.~\ref{Fig:FuhrerA} we show our experimental data supporting a
percolation driven metal-insulator transition on a naturally occurring
graphene nanoribbon (that is expected to have smoother edges than
comparable ribbons fabricated using the method of
Ref.~\cite{kn:han2007}) of dimensions $L \approx 1.5~{\mu m}$ and $W
\approx 50~{nm}$.  The conductance $G = (W/L) \sigma$ is shown as a
function of applied gate voltage $V_g \propto n$.  We performed
Quantum Hall measurements on the large area connected to the ribbons
to confirm that we have a mono-layer of graphene and the details of
the fabrication can be found in Ref.~\cite{kn:chen2008}.  Best fits to
Eq.~\ref{Eq:Percolation} at low density give conductivity exponents
$\delta^e = 1.3 \pm 0.2$ and $\delta^h = 1.3 \pm 0.1$ (close to the
theoretically expected value $\delta = 4/3$), and the fit
to Eq.~\ref{Eq:Boltzmann} at high density gives $n_{\rm imp} = 22
\times 10^{10}~{\rm cm}^{-2}$ which is consistent with measurements on
similarly prepared bulk graphene samples~\cite{kn:chen2008}.
Fig.~\ref{Fig:PanelFigure} shows a similar analysis for a $W=200~{\rm
nm}$ sample fabricated at Maryland, and two of the Columbia samples
($W = 24~{\rm nm}$ and $W = 49~{\rm nm}$) reported in
Ref.~\cite{kn:han2007}.  All three samples show the low-density
percolation universality class with critical exponents $\delta = 1.2
\pm 0.2,1.3 \pm 0.1$, and $1.55 \pm 0.3$ respectively, which are
similar to percolation exponents observed in 2D GaAs
systems~\cite{kn:dassarma2005b,kn:manfra2007}.  The corresponding fit
parameters for the three samples are $W A/L = 0.008,0.002$ and
$0.0087$ respectively, and $n_c = -65.08, 170.87$ and
$184.23~\times~10^{10}~{\rm cm}^{-2}$ respectively (the units of $A$
are $[(10^{10}~{\rm cm}^{-2})^{\delta}~25.8~k\Omega]^{-1}$).  

For the 2D percolation universality class, at low density we have
\begin{equation} 
\sigma = A (n - n_c)^\delta
\label{Eq:Percolation}
\end{equation}
where $\delta \approx 4/3$ is the 2D percolation critical exponent.
For graphene ribbons, we expect two such percolation transitions, one
for electrons and one for holes, separated by a ``transport gap''
defined as $\Delta_g = \gamma \sqrt{ \pi (n_c^e - n_c^h)}$, where
$n_c^{e(h)}$ is the critical density for electrons (holes) and $\gamma
= \hbar v_{\rm F}$ is the graphene Fermi velocity.  For larger carrier
densities, where $|E_{\rm F}| \gg \Delta_g$, we expect a crossover to
a high-density Boltzmann transport regime where~\cite{kn:hwang2006c,kn:adam2007a}
\begin{equation}
\sigma = 20 \frac{e^2}{h} \left(\frac{|n - n_{\rm D}|}{n_{\rm imp}}\right),
\label{Eq:Boltzmann}
\end{equation}
just as for bulk graphene on a SiO$_2$ substrate (where $n_{\rm D}$ 
is the charge neutrality point and $n_{\rm imp}$ is the $2D$ surface
impurity density of Coulomb scatterers). If we define $\xi$
as the typical size of the electron or hole puddle, where below
we calculate $\xi$ self-consistently using the random phase
screening approximation, then so long as the sample width $W \gtrsim
\xi$, we would have 2D percolation whereas if $W \lesssim \xi$ one
has 1D percolation i.e. a chain of approximately $L/\xi$ p-n
junctions.  Changing $n_{\rm imp}$ (which could be extracted from
high density mobility measurements) would also change $\xi$ and the 
critical width for which this dimensional
crossover is observed.  Moreover, we predict that signatures of reduced 
dimensionality should be apparent in temperature dependent 
transport measurements.

To date, most theories for transport in graphene nanoribbons consider
a quasi 1D, rather than the 2D limit.  The experimental observation of 2D
percolation in these ribbons casts strong doubt onto the relevance of
the quasi 1D theories for current graphene nanoribbon experiments.
As was already discussed in Ref.~\cite{kn:nakada1996}, many features
of the quasi 1D geometry get washed out for $W \gtrsim 10~{\rm nm}$
which is the case in most experiments on graphene nanoribbons.
The length scale controlling the crossover from quasi 1D to 2D
behavior in this context may very well be determined by other
independent parameters which are unknown at this stage such as the
inelastic scattering length or the phase breaking length (both of
which depend on temperature).  Whether the transport properties in
graphene nanoribbons should be considered using a 2D or a quasi 1D
Hamiltonian is at this stage an open question requiring further
theoretical and experimental study.  Our analyses involving data from 
two groups, our own and that of the Columbia group~\cite{kn:han2007}, 
clearly establish that depending on the value of $W$ either the 2D or 1D 
percolation universality class may apply,
where we believe this transition to be controlled by the size of the
electron and hole puddles induced by charged impurities.  We can not rule
out the possibility that further lowering of temperature
would lead to quasi 1D behavior~\cite{kn:sols2007} and the
percolation-driven 2D MIT is only a crossover phenomenon.  Although we
focus on single-layer graphene, we note that a similar
percolation transition should also be seen in graphene bilayers, where
since an electric field induced gap can be introduced into the
spectrum without any confinement, the crossover to a quasi 1D regime
would not arise.  We note that even for bulk graphene, a
Boltzmann to percolation crossover could be induced with a magnetic
field, where for small field and within the electron-hole puddle
model, we expect the p-n resistance to be very low justifying the
Boltzmann picture, whereas for large magnetic field, the p-n junction
becomes very resistive~\cite{kn:cheianov2006b} inducing a percolation
transition.  This crossover may have been observed in recent
experiments~\cite{kn:checkelsky2008}.

To reinforce the point that the 2D MIT in graphene nanoribbons is
indeed a percolation transition and {\it not} a quantum crossover
phenomenon, we calculate the percolation critical density $n_p$ using
the non-linear screening argument of Efros~\cite{kn:efros1988} with
the basic idea being that the MIT occurs when inhomogeneous density fluctuations
created by the charged impurities can no longer be screened by the
carriers.  This leads to $n_p \sim \sqrt{n_{\rm imp}}/d$ where the
random charged impurities of concentration $n_{\rm imp}$ are assumed
to be located at a distance $d$ from the 2D graphene plane.  Taking $d
\sim 1~{\rm nm}$ and $n_{\rm imp} \sim 2-5 \times 10^{11}~{\rm
cm}^{-2}$, typical values
estimated~\cite{kn:hwang2006c,kn:adam2007a,kn:tan2007,kn:chen2008}
from mobility measurements, we get $n_p \sim 5 \times 10^{12}~{\rm
cm}^{-2}$.  This is in reasonable agreement with our experimental
finding in Figs.~\ref{Fig:FuhrerA} and
~\ref{Fig:PanelFigure}.  On the other hand, the quantum localization
crossover density $n_q$ can be estimated from the Ioffe-Regel
criterion $k_{\rm F} \ell \sim 1$, where $\ell$ is the mean-free path,
to be $n_q \sim 2\times 10^{10} {\rm cm}^{-2}$ for the same $n_i$ and
$d$ values.  Thus, $n_q \ll n_p$, and our experimental critical
density agrees with the percolation critical density, providing
further support for a percolation driven insulating transition in
graphene.

Experimentally, one can measure three different gaps.  In addition to
the transport gap $\Delta_g$ discussed above, the temperature
dependence of the conductivity minimum gives an activated gap
$\Delta_{\rm act}$ (we performed this measurement on the $200~{\rm
nm}$ ribbon, and found $\Delta_{\rm act} W \approx 0.1~{\rm eV~nm}$
which is an order of magnitude smaller than theoretical
estimates~\cite{kn:yang2007} of the confinement induced
gap), and finally, Ref.~\cite{kn:han2007} reported the source-drain
bias required to induce conduction and found the gap to be orientation
independent (contrary to the expectation of the quasi 1D theory that
is extremely sensitive to whether the edge is zig-zag or armchair).
The connection between these three experimental gaps and the
theoretical gap in the energy spectrum is beyond the scope of this
work, where we focus here only on the transport gap $\Delta_g$ and
predict that in the 2D regime, provided that the impurity location is
pretty much the same, $|n_c^e - n_c^h| \propto \sqrt{n_{\rm imp}}
\propto \mu^{-1/2}$, where $\mu$ is the high-density mobility in the
Boltzmann
regime~\cite{kn:hwang2006c,kn:adam2007a,kn:tan2007,kn:chen2008}.
 
The 2D percolation picture presented above breaks down when the 
sample width becomes smaller than the typical disorder length scale.
Using the self-consistent RPA method of Ref.~\cite{kn:adam2007a} we 
can obtain an integral expression for the potential correlation 
function $\langle V(r) V(0) \rangle$ which for experimentally
relevant parameters can be approximated by
\begin{equation}
\langle V(r) V(0) \rangle \approx 
\frac{K_0 \gamma^2}{2 \pi \xi^2} \exp\left[\frac{-r^2}{2 \xi^2}\right].
\end{equation}
Using $r_{s} = e^2/\kappa \gamma$, where $\kappa$ is the effective dielectric
constant that depends on the choice of substrate, we find 
\begin{subequations}
\begin{eqnarray}
K_0 &=& \frac{1}{4 r_s^2} \left( \frac{D_0}{C_0} \right)^2, \\
\xi &=& \frac{1}{\sqrt{n_{\rm imp}}} \frac{D_0}{4 \pi r_s^2} 
\frac{1}{(C_0)^{3/2}},
\end{eqnarray}
 \label{Eq:Mapping}
\end{subequations}  
where for $z = 4 k_{\rm F} d$ and 
$E_1(x) = \int_x^\infty t^{-1} e^{-t} dt$     
\begin{widetext}
\begin{subequations}
\begin{eqnarray}
C_0(z) &=& -1 + \frac{4 E_1(z)}{(2 + \pi r_s)^2} + 
        \frac{2 e^{-z} r_s}{1 + 2 r_s} 
         + (1+2 z r_s)e^{2 z r_s}(E_1[2 z r_s] - E_1[z(1+2 r_s)]),   \\
D_0(z) &=& 1 - \frac{8 r_s z E_1[z]}{(2 + \pi r_s)^2} 
             + \frac{8 e^{-z} r_s}{(2 + \pi r_s)^2} 
             - \frac{2 e^{-z} r_s}{1+ 2 r_s}
             - 2 z r_s e^{2 z r_s}(E_1[2 z r_s] - E_1[z(1+ 2 r_s)]).
\end{eqnarray}
\end{subequations}
\end{widetext}
This notation is chosen to be consistent with Ref.~\cite{kn:adam2007a}
where the {\it rms} density $n^* = 2 r_s^2 n_{\rm imp} C_0(z = 4 d
\sqrt{\pi n^*})$.  For typical values of $n_{\rm imp} \approx 20\times
10^{10}~{\rm cm}^{-2}$ and $d \approx 1~{\rm nm}$, we have $\xi
\approx 10~{\rm nm}$, which is consistent with the experimentally
observed critical width $W^* = 16~{\rm nm}$~\cite{kn:han2007}.  In the
1D limit, there should be no percolation transition, only an activated
conduction, and the effective gap should diverge in the $T\rightarrow
0$ limit.  From Eq.~\ref{Eq:Mapping}, we can predict the dependence of
the puddle size (and therefore the critical width of the dimensional
crossover) on experimentally tunable parameters.  For example, we
predict that cleaner samples (i.e. with a larger high-density
mobility) would have {\it larger} critical widths and by doping
graphene with potassium~\cite{kn:chen2008}, thereby changing only
$n_{\rm imp}$, one could tune through this dimensional crossover in a
sample of fixed width.  Changing the substrate to a high-$\kappa$
material like HfO$_2$ (assuming that $n_{\rm imp}$ and $d$ remain
unchanged) could significantly increase the puddle size, in contrast
to suspended graphene~\cite{kn:bolotin2008} where
increasing $r_s$ (which decreases the puddle size) is compensated by
the lower $n_{\rm imp}$ (which tends to increase the puddle size).  In
addition, we predict that suspended nanoribbon experiments will have a
smaller transport gap due to the order of magnitude higher mobility,
but the same critical exponent $\delta$; while for nanoribbons with
potassium doping of various strengths, increasing $n_{\rm imp}$ should
lead to a larger transport gap with no change in $\delta$.  

In conclusion, we have argued theoretically and demonstrated
experimentally that a disorder induced, density inhomogeneity driven
percolation transition is observable in graphene nanoribbons.  We
anticipate a crossover to Boltzmann transport at high carrier density
and a dimensional crossover for sample widths that are smaller than
the disorder induced puddle size.  Several features of the experiment
including the difference between transport and activation gaps, the
large discrepancy between the value of the gap and that predicted by
band structure theory as well as there being no orientation dependence
of the gap and a critical width below which the dimensional crossover
causes a divergence in the transport gap (for an infinite 1D system 
in the $T\rightarrow 0$ limit) are all explained naturally in this
picture.  This consistent theoretical picture better captures the
physics of nanoribbons than the quasi 1D induced models that have
dominated the literature to date.  Our discovery
of a percolation driven graphene 2D MIT also shows the close conceptual
connection between 2D graphene transport and 2D semiconductor
transport, and establishes that density inhomogeneities
dominate carrier transport in both classes of systems at low enough
carrier densities.  In bulk 2D graphene, which is a zero-gap
semiconductor, this leads to the low-density minimum conductivity
plateau, and in graphene nanoribbons (as well as in bilayer graphene
with an electric field induced gap), where there is an energy gap
between the electron and hole bands, we get the same percolation
induced 2D MIT familiar from 2D semiconductor
electron~\cite{kn:dassarma2005b} and hole~\cite{kn:manfra2007}
systems.  For very narrow graphene nanoribbons, which are in the 1D
percolation regime, our theory predicts an insulating behavior with an
effective infinite energy transport gap at $T=0$, which may have been
observed experimentally~\cite{kn:han2007,kn:li2008}, by virtue of the
absence of a percolation transition in 1D.  A fundamental question of
considerable significance that remains open in this context is the
experimental absence of quantum localization~\cite{kn:aleiner2006}
which may be observable at much lower temperatures than used
experimentally so far.

This work is partially supported by U.S. ONR. 


\end{document}